\documentclass[10pt]{iopart}
\usepackage{graphicx}

\newcommand{\lf}[2]{\mbox{\Large $\frac{#1}{#2}$}}

\begin{document}

%

\jl{6}

\title[Fake signals\ldots]{Fake signals caused by heavy mass motions
near a sensitive spherical gravitational wave antenna}

\author{Alberto Lobo\dag\footnote[3]{To whom correspondence should be
addressed.}, Massimo Cerdonio\ddag and \'Alvaro Montero\dag}

\address{\dag\ Departament de F\'\i sica Fonamental, Universitat de
	Barcelona, Diagonal 647, 08028 Barcelona, Spain.}

\address{\ddag\ INFN, Padova Section, and Department of Physics,
	Universit\'a di Padova, via Marzolo 8, I-35100 Padova, Italy}

\date{\today}

\begin{abstract}
This paper analyses in quantitative detail the effect caused by a
moving mass on a spherical gravitational wave detector. This applies
to situations where heavy traffic or similar disturbances happen near
the GW antenna. Such disturbances result in quadrupole \emph{tidal}
stresses in the antenna mass, and they therefore precisely fake
a real gravitational signal. The study shows that there always are
characteristic frequencies, depending on the motion of the external
masses, at which the fake signals are most intense. It however appears
that, even at those frequencies, fake signals should be orders of
magnitude below the sensitivity curve of an optimised detector, in
likely realistic situations.
\end{abstract}




\section{Introduction
\label{sec.1}}

Large mass spherical gravitational wave (GW) detectors \cite{lobo,clo}
constitute sensitive systems with a real promise of sighting GW events
in the frequency range of 1 kHz, even with a rather large bandwidth
\cite{dual}. In such extremely delicate device as an acoustic GW
detector is~\cite{as93,alle,niobe,naut,aur}, any disturbance, whether
internal or external, must be considered and, if possible, screened out
both in hardware and in software~\cite{chi2,igec}. Quite recently, a
possibility has arisen to build an underground spherical detector in
the Spanish Pyrinees~\cite{mora}, located near a road tunnel.

Heavy traffic in the tunnel is there almost at all times. It is therefore
important to assess the effect of the passing vehicles close to the
antenna. Purely Newtonian action is expected on the sphere which exactly
fakes a real GW signal, since it shows up as a local \emph{tide}, with
its characteristic quadrupole structure. Hope is that such fake signals
be weak compared to actual GW signals. Otherwise, even if veto control
on them could easily be exercised, an almost continuous background of
spurious signals would contaminate the detector output, eventually
concealing a real signal.

The purpose of this paper is to calculate the effect of traffic on the
spherical detector by means of a point moving mass model, and to present
the results of applying the theory to a realistic situation. We shall
first consider the effect of a mass moving with \emph{uniform speed}
past the detector. As we shall see, this causes a very small impact
on the detector in any reasonable circumstances, and we thus consider
next the effect of oscillating parts in the passing vehicle, which can
be more important. In section~\ref{sec.2} we pose the general problem,
then in sections~\ref{sec.3} and~\ref{sec.4} we fully consider each
case; section~\ref{sec.5} is finally devoted to comment on the results
obtained and their practical relevance in a real detector site.

\section{The problem
\label{sec.2}}

\begin{figure}[t]
\centering
\includegraphics[width=10cm]{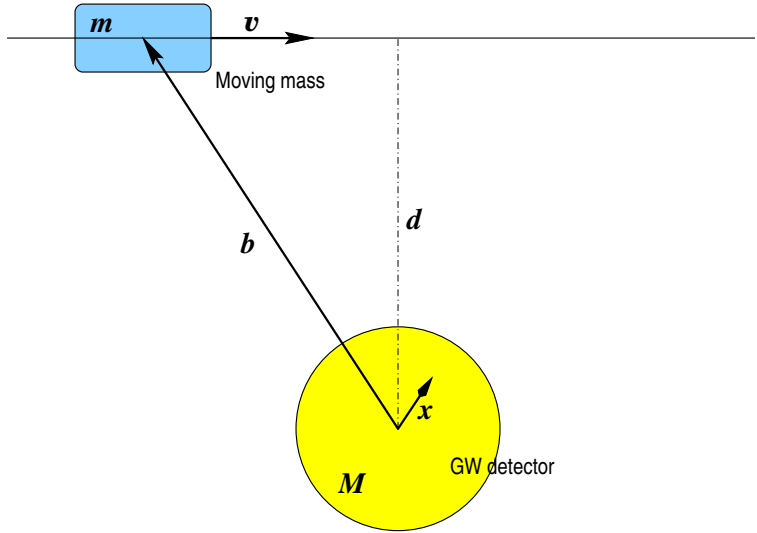}
\caption{Schematics of a passing mass and a spherical GW detector.
Distances do not match the scales of any realistic example.
\label{fig.1}}
\end{figure}

The situation is best described by the schematic graphic displayed
in figure~\ref{fig.1}: a spherical GW detector has its centre at the
origin of coordinates, and a moving mass~$m\/$ travels in a straight
line past the detector with a given \emph{impact parameter} $|\bi{d}|$.
We shall assume that the detector's radius $R\/$ is much smaller than
this parameter, i.e.,

\begin{equation}
 R\ll |{\bi{d}}|\;,
 \label{eq.1}
\end{equation}
and ask which is the Newtonian \emph{gravitational action} of the moving
mass on the GW detector.

Let then $\phi(\bi{x},t)$ be the \emph{gravitational potential}
generated by the moving mass at point $\bi{x}$ at time~$t$. We
shall of course assume it satisfies Poisson's equation

\begin{equation}
 \nabla^2\phi(\bi{x},t) = -4\pi G\,\rho(\bi{x},t)
 \label{eq.2}
\end{equation}
where $G\/$ is Newton's constant, and $\rho(\bi{x},t)$ is the mass
density distribution of the moving mass. We shall also assume this
sufficiently small that it can be considered a \emph{point mass}, i.e.,

\begin{equation}
 \rho(\bi{x},t) = m\,\delta^{(3)}(\bi{x}-\bi{b})
 \label{eq.3}
\end{equation}
where (see again figure~\ref{fig.1})

\begin{equation}
 \bi{b} = \bi{b}(t) = \bi{d} + \bi{v}t
 \label{eq.4}
\end{equation}
if one adopts the (arbitrary) convention that the moving mass crosses
the point of closest distance to the detector, $\bi{d}$, at time $t=0$,
and that it moves with constant velocity $\bi{v}$. The gravitational
potential is thus

\begin{equation}
 \phi(\bi{x},t) = -\frac{Gm}{|\bi{x}-\bi{b}(t)|}
 \label{eq.5}
\end{equation}

GWs show up in the detector as local \emph{tides}, therefore the
moving mass will cause signal confusion inasmuch as it too produces
tides. Under the assumption~(\ref{eq.1}) above, we can calculate the
gravitational potential \emph{inside} the sphere by the approximate
expansion

\begin{equation}
 \phi(\bi{x},t) = \phi({\bf 0},t) + \partial_i\phi({\bf 0},t)\,x_i +
 \lf{1}{2}\;\partial_i\partial_j\phi({\bf 0},t)\,x_ix_j + \ldots
 \label{eq.6}
\end{equation}

The gravitational \emph{accelerations} are correspondingly given by
the gradient of the potential, i.e., $\bi{a}$\,=\,$-\nabla\phi$, or

\begin{equation}
 a_i(\bi{x},t) = -\partial_i\phi({\bf 0},t)\,x_i
 - \partial_i\partial_j\phi({\bf 0},t)\,x_j + \ldots
 \label{eq.7}
\end{equation}

The first term in the rhs represents the \emph{global pull} on the
sphere, and cannot therefore excite its oscillation modes; these are
instead excited by the \emph{acceleration gradients}, or tides, given
by the second term. These acceleration gradients can be converted to a
\emph{density of forces} (force per unit volume) simply by multiplying
them by the density of the sphere,~$\varrho$, say, to obtain

\begin{equation}
 f_i^{\rm MM}(\bi{x},t) = -\varrho\;
 \partial_i\partial_j\phi({\bf 0},t)\,x_j\equiv
 \varrho\,R_{ij}^{\rm MM}(t)\,x_j
 \label{eq.8}
\end{equation}

We readily find

\begin{equation}
 R_{ij}^{\rm MM}(t) = -\frac{3Gm}{b^3(t)}\;\left(
 \frac{b_i(t)\,b_j(t)}{b^2(t)}-\frac{1}{3}\,\delta_{ij}\right)
 \label{eq.9}
\end{equation}
where $b(t)\equiv|\bi{b}(t)|$. The interaction of the passing mass
with the sphere is described by the usual equations of elasticity
theory \cite{ll70}

\begin{equation}
 \varrho\,\frac{\partial^2\bi{u}}{\partial t^2}-\mu\nabla^2 \bi{u}
 -(\lambda+\mu)\,\nabla(\nabla{\bf\cdot}\bi{u}) = \bi{f}(\bi{x},t)\;.
 \label{eq.10}
\end{equation}

The structure of $\bi{f}^{\rm MM}(\bi{x},t)$ is obviously purely
quadrupole, so it can be split up much in the same way as a GW
excitation\footnote{
Notation and conventions on sphere's modes and response will follow
closely those of reference~\protect\cite{lobo} henceforth.}:

\begin{equation}
 \bi{f}^{\rm MM}(\bi{x},t) =
 \sum_{m=-2}^2\,\bi{f}^{(m)}(\bi{x})\,C^{(m)}(t)
 \label{eq.11}
\end{equation}
where

\begin{equation}
 f_i^{(m)}(\bi{x}) = \varrho\,E_{ij}^{(2m)}\,x_j\ ,\qquad
 C^{(m)}(t) = \frac{8\pi}{15}\,E_{ij}^{*(2m)}\,
 R_{ij}^{\rm MM}(t)
 \label{eq.12}
\end{equation}
with $m = -2,\ldots,2$. The matrices $E_{ij}^{(2m)}$ are given
by~\cite{lobo}

\numparts
\begin{eqnarray}
 E_{ij}^{(20)} & = & \left(\mbox{\large $\frac{5}{16\pi}$}
                \right)^{\!\frac{1}{2}}\left[\begin{array}{ccc}
  -1 & 0 & 0 \\ 0 & -1 & 0 \\ 0 & 0 & 2 \end{array}\right] \\[1 ex]
 E_{ij}^{(2\pm 1)} & = & \left(\mbox{\large $\frac{15}{32\pi}$}
                \right)^{\!\frac{1}{2}}\left[\begin{array}{ccc}
   0 & 0 & \mp 1 \\ 0 & 0 & -i \\ \mp 1 & -i & 0 \end{array}\right] \\[1 ex]
 E_{ij}^{(2\pm 2)} & = & \left(\mbox{\large $\frac{15}{32\pi}$}
                 \right)^{\!\frac{1}{2}}\left[\begin{array}{ccc}
   1 & \pm i & 0 \\ \pm i & -1 & 0 \\ 0 & 0 & 0 \end{array}\right]\;.
 \label{eq.125}
\end{eqnarray}
\endnumparts

The sphere's response to the forces (\ref{eq.11}) is thus determined
by the series expansion

\begin{equation}
 \bi{u}^{\rm MM}(\bi{x},t) = \sum_{n=1}^\infty\,
 \frac{a_{n2}}{\omega_{n2}}\,\left[
 \sum_{m=-2}^2\,\bi{u}_{n2m}(\bi{x})\,C_{n2}^{(m)}(t)\right]
 \label{eq.13}
\end{equation}
which is formally identical to the response to a GW excitation \footnote{
Here, $C_{n2}^{(m)}(t)$ is a convolution product
\[
 C_{n2}^{(m)}(t) = \int_0^t\,C^{(m)}(t')\,\sin\omega_{n2}(t-t')\,dt'
\]
between the driving term, $C^{(m)}(t)$, and the antenna mode,
$\sin\omega_{n2}(t)$. In a real system, the latter has an additional
damping factor, $e^{-\gamma_{n2}t}$, which gives a system transfer
function with a non-zero linewidth, $\gamma_{n2}$.}.
Our concern is now to compare this with the system's response to
a \emph{real} GW. We come to it in the next section.

\section{Equivalent signal
\label{sec.3}}

A meaningful comparison between the passing mass fake signal and an
actual GW signal is best set up in \emph{frequency domain}, as it is
in this form that the detector's sensitivity is defined.

Consider first a point on the sphere's surface, $\bi{x}=R\bi{n}$,
where $\bi{n}$ is a unit outward pointing normal, and calculate
its \emph{radial} displacement $u(t)$, i.e.,

\begin{equation}
 u(t)\equiv\bi{n}\cdot\bi{u}(R\bi{n},t)\;.
 \label{eq.3-1}
\end{equation}

Because of the form of the sphere's wavefunctions, the Fourier transform
of the above is easily seen to be~\cite{dual}

\begin{equation}
 \tilde u^{\rm GW}(\omega) = -\frac{1}{2}\sum_{n=0}^\infty a_{n2}A_{n2}(R)\,
 \omega^2\tilde{h}(\omega)\,L_{n2}(\omega)\;,
 \label{eq.3-3}
\end{equation}
where $L_{n2}(\omega)$ is the mode's transfer function

\begin{equation}
 L_{n2}(\omega) = \frac{1}{-\omega^2+2i\gamma_{n2}\;\omega+\omega_{n2}^2}
 \label{eq.3-4}
\end{equation}
with $\gamma_{n2}$ the linewidth of the mode, and $\tilde{h}(\omega)$ the
Fourier transform of the signal $h(t)$, defined by

\begin{equation}
 h(t)\equiv h_{ij}(t)\,n_in_j
 \label{eq.3-5}
\end{equation}

Likewise, the radial displacement induced at the same location of the
sphere by the moving mass is easily inferred from equation~(\ref{eq.13}):

\begin{equation}
 \tilde u^{\rm MM}(\omega) = \sum_{n=0}^\infty a_{n2}A_{n2}(R)\,
 \tilde{R}_{ij}^{\rm MM}(\omega)\,n_in_j\,L_{n2}(\omega)
 \label{eq.3-6}
\end{equation}

We can at this point define an \emph{equivalent} amplitude associated
to the fake moving mass signal; the natural definition~is~\cite{dual}:

\begin{equation}
 \left|\tilde{h}_{\rm equiv}^{\rm MM}(\omega)\right|^2\equiv\frac
 {\left|\tilde{u}^{\rm MM}(\omega)\right|^2}
 {\left|\tilde{u}^{\rm GW}(\omega)\right|^2/|\tilde{h}(\omega)|^2}
 \label{eq.3-7}
\end{equation}

We thus find

\begin{equation}
 \left|\tilde{h}_{\rm equiv}^{\rm MM}(\omega)\right|^2 =
 4\omega^{-4}\,\left|\tilde{R}_{ij}^{\rm MM}(\omega)\,n_in_j\right|^2\;.
 \label{eq.3-8}
\end{equation}

It is important to stress here that this expression holds for \emph{solid}
spheres as well as for \emph{hollow} spheres, and indeed for a \emph{dual}
sphere, too. This is simply because the \emph{mode expansion} is identical
in the GW signal and in the passing mass signal.

An order of magnitude estimate of the passing mass equivalent
signal is accomplished by replacing
$\left|\tilde{R}_{ij}^{\rm MM}(\omega)\,n_in_j\right|^2$
in equation~(\ref{eq.3-8}) with
$\left[\tilde{R}_{ij}^{*{\rm MM}}(\omega)
\tilde{R}_{ij}^{\rm MM}(\omega)\right]$, i.e.,

\begin{equation}
 \left|\tilde{h}_{\rm equiv}^{\rm MM}(\omega)\right|^2\leq
 4\omega^{-4}\;\tilde{R}^{*{\rm MM}}_{ij}(\omega)
 \tilde{R}_{ij}^{\rm MM}(\omega)\;,
 \label{eq.3-9}
\end{equation}
%

The Fourier transforms $\tilde{R}_{ij}^{\rm MM}(\omega)$ are expediently
calculated adopting a coordinate system where the $x\/$-axis is
parallel to the velocity vector $\bi{v}$, and the $y\/$-axis is
in the direction of the vector $\bi{b}$ of figure~\ref{fig.1}.
The result is

\begin{equation}
 \hspace*{-2 cm}
 \tilde R_{ij}^{\rm MM}(\omega) =
 \frac{Gm}{vd^2}\,\left(
 \begin{array}{ccc}
 z^2K_2(z)-2zK_1(z) & -iz^2K_1(z) & 0 \\
 -iz^2K_1(z) & -z^2K_2(z)+zK_1(z) & 0 \\
 0 & 0 & zK_1(z)
 \end{array}\right)
 \label{eq.19}
\end{equation}
where $d\equiv|\bi{d}|$, and

\begin{equation}
 z\equiv\omega/\omega_0\ ,\qquad
 \omega_0\equiv\frac{v}{d}
 \label{eq.20}
\end{equation}
with the $K\/$'s representing modified Bessel functions \cite{abra}:

\begin{equation}
 K_\nu(z) = \frac{\Gamma\left(\nu+1/2\right)}
 {\Gamma\left(1/2\right)}\,\left(\frac{2}{z}\right)^{\!\nu}\;
 \int_0^\infty\;\frac{\cos zt}{(1+t^2)^{\nu+1/2}}\,dt\;.
 \label{eq.22}
\end{equation}

It is now easily seen that

\begin{equation}
 \left|\tilde{h}_{\rm equiv}^{\rm MM}(\omega)\right|\leq
 \frac{Gm}{vd^2}\;\omega^{-2}\,f(z)
 \label{eq.23}
\end{equation}
where

\begin{equation}
 f(z)\equiv\left[
 z^4K_2^2(z) + (3+z^2)\,z^2K_1^2(z) - 3z^3K_1(z)K_2(z)\right]^{1/2}\;.
 \label{eq.24}
\end{equation}

\begin{figure}[t]
\centering
\includegraphics[width=11cm]{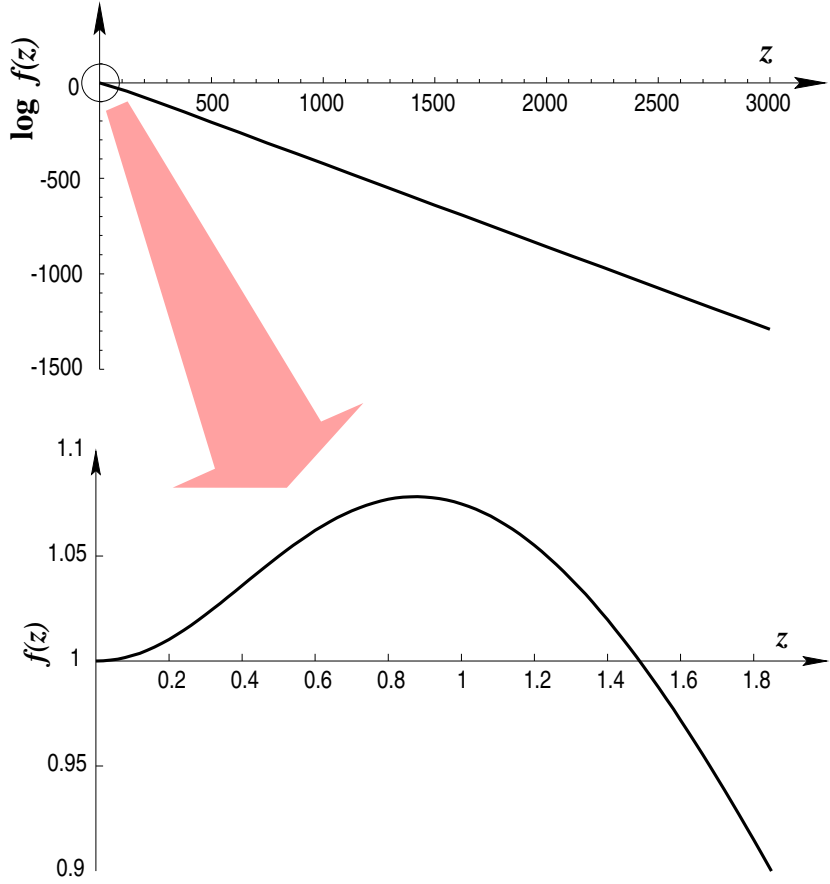}
\caption{Above: the spectral density function $f(z)$ of equation
(\protect\ref{eq.24}), plotted on a logarithmic scale. Below: close-up
of the above in the low frequency range, showing the function's maximum
at $z\simeq 0.88$. Linear scales have been used here for clarity.
\label{fig.2}}
\end{figure}

This function has asymptotic behaviours:

\begin{eqnarray}
 f(z)& \sim & 1 + \left(\frac{z}{2}\right)^{\!2}
 \ ,\ \ \qquad {\rm if}\ \ z\ll 1
 \label{eq.24a} \\[1 ex]
 f(z)& \sim & \sqrt{\pi}\,z^{3/2}\,e^{-z}\ ,\qquad {\rm if}\ \ z\gg 1\;,
 \label{eq.24b}
\end{eqnarray}
and has a rather smooth maximum at $z\sim 1$, as we see in the plot of
figure~\ref{fig.2}.

The salient feature of $f(z)$ as regards our present concern is however
its exponential roll-off in the high frequency range. Take for example as
reasonable figures a somewhat fast 10 ton truck travelling at a speed of
40 m/s ($\sim$\,140 km/h), with an \emph{impact parameter} $d$\,=\,80 m;
this gives a frequency $v/d$\,=\,0.5 rad/sec, or $\sim$\,0.1 Hz, which
means we must go up to $z\sim 10^4$ for frequencies in the kHz range, where
a spherical detector will be sensitive. But this produces an utterly
meaningless $\tilde{h}_{\rm equiv}^{\rm MM}\sim 10^{-4354}$ at
$\omega/2\pi$\,=\,1 kHz \ldots

Such number clearly indicates that anything will produce more substantial
fake signals in the detector, provided it has some vigor in the kHz range.
An example could be e.g.\ internal motions of parts of the vehicle, such
as the engine pistons, which undergo periodic oscillations of a more
appreciable magnitude. We come to this next.

\section{Periodic source
\label{sec.4}}

\begin{figure}[t]
\centering
\includegraphics[width=10cm]{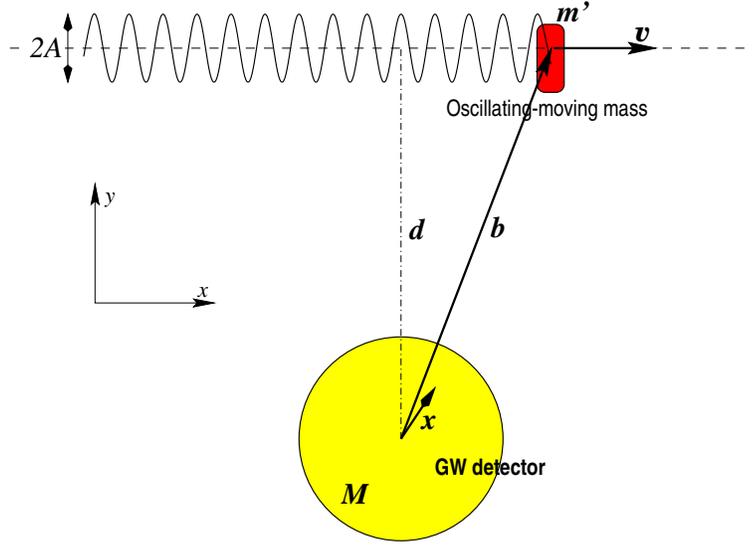}
\caption{Schematics of a spherical GW detector and a passing mass which,
simultaneously, oscillates with amplitude $A$. Again, distances do not
match realistic scales.
\label{fig.3}}
\end{figure}

Consider thus that certain internal parts of the passing vehicle
oscillate with (angular) frequency $\alpha$ and amplitude~$A$ ---see
figure~\ref{fig.3}. If we represent with $m'$ the mass of one such part
then the Newtonian potential at a field position $\bi{x}$ relative to
the detector's centre is obviously given by

\begin{equation}
 \phi(\bi{x},t) = -\frac{Gm'}{|\bi{x}-\bi{b}(t)-\bi{c}(t)|}
 \label{eq.4-1}
\end{equation}
where

\begin{equation}
 \bi{b}(t) = (vt,0,0)\ ,\qquad
 \bi{c}(t) = (0,A\cos\alpha t,0)
 \label{eq.4-2}
\end{equation}
with the axes convention shown in figure~\ref{fig.3}. The system
response will be given in this case by the same formulas of
section~\ref{sec.2}, except that the quadrupole matrix
$R_{ij}^{\rm MM}(\omega)$ needs to be replaced with

\begin{equation}
 \hspace*{-1 cm}
 R_{ij}^{\rm OMM}(t) = -\frac{3Gm'}{|\bi{b}(t)+\bi{c}(t)|^3}\;\left(
 \frac{[b_i(t)+c_i(t)]\,[b_j(t)+c_j(t)]}{|\bi{b}(t)+\bi{c}(t)|^2}
 -\frac{1}{3}\,\delta_{ij}\right)
 \label{eq.4-3}
\end{equation}
and, in particular, the new \emph{equivalent signal} will be given by

\begin{equation}
 \left|\tilde{h}_{\rm equiv}^{\rm OMM}(\omega)\right|^2\leq
 4\omega^{-4}\;\tilde{R}^{*{\rm OMM}}_{ij}(\omega)
 \tilde{R}_{ij}^{\rm OMM}(\omega)\;,
 \label{eq.4-4}
\end{equation}
by the same arguments which led to equation~(\ref{eq.3-9}).

The Fourier transforms of the functions $R_{ij}^{\rm OMM}(t)$ in
equation~(\ref{eq.4-3}) are too complicated to be performed
analytically, but there are certain features which can be easily
inferred by inspection of the defining formulas. Take for example
the gravitational potential given by equation~(\ref{eq.4-1}); we
can recast this in the form

\begin{equation}
 \phi(\bi{x},t) = -\frac{Gm'}{|\bi{x}-\bi{b}|}\,\left[
  1 - \frac{2\bi{c}\cdot(\bi{x}-\bi{b})}{|\bi{x}-\bi{b}|^2}
  + \frac{|\bi{c}|^2}{|\bi{x}-\bi{b}|^2}\right]^{-1/2}
 \label{eq.4-5}
\end{equation}
and perform a power series expansion of the term in square brackets
in the rhs on the basis that

\begin{equation}
 |\bi{c}|\ll |\bi{x}-\bi{b}|
 \label{eq.4-6}
\end{equation}
i.e., actually assuming that the amplitude $A$ of oscillations of the
mass~$m'$ is much smaller than the \emph{impact parameter} $|\bi{d}|$
---see figure~\ref{fig.3}. The series expansion will thus consist in
a series of ascending powers of $|\bi{c}|=A\cos\alpha t$, so it will
consequently involve all the \emph{harmonics} of the frequency
$\alpha/2\pi$, clearly with an amplitude which decreases with the
harmonic order due to the inequality~(\ref{eq.4-6}). Clearly, the
same harmonic structure carries over to the higher derivatives of
the potential, and in particular to the quadrupole moments
$R_{ij}^{\rm OMM}(t)$.

Since we must give up the analytic approach of section~\ref{sec.3}, we
shall now estimate the Fourier transforms $\tilde{R}_{ij}^{\rm OMM}(\omega)$
by numerical methods. For this we shall make the rather natural choice of
using \emph{Fast Fourier Transform} algorithms~\cite{fft}.

In this approach, first thing we need is to define an appropriate
\emph{bandwidth} to do the analysis. Our standard reference will be
a \emph{dual sphere} GW detector, which is the best spherical GW
antenna we can think of at present ---see~\cite{dual}. To accurately
match that reference we must assess the signal intensities up to a
frequency of 3000 Hz, hence we need a bandwidth of 6000 Hz, including
of course negative frequencies. We shall thus approximate Fourier
transforms by DFT sums:

\begin{equation}
 \hspace*{-1.5 cm}
 \tilde{R}_{ij}^{\rm OMM}(\omega_n) = \Delta t\;
 \sum_{m=0}^{N-1}\,R_{ij}^{\rm OMM}(m\,\Delta t)\,e^{2\pi inm/N}\ ,
 \qquad \omega_n = n\Omega_{\rm Nyq}/N\;,
 \label{eq.4-7}
\end{equation}
where $N\/$ is the total number of sample points, $\Omega_{\rm Nyq}$
is the Nyquist, or sampling frequency, and $\Delta t$ is the time
interval between successive samples, i.e.,

\begin{equation}
 \Delta t = \frac{2\pi}{\Omega_{\rm Nyq}}\;.
 \label{eq.4-8}
\end{equation}

\begin{figure}[b]
\centering
\includegraphics[width=12cm]{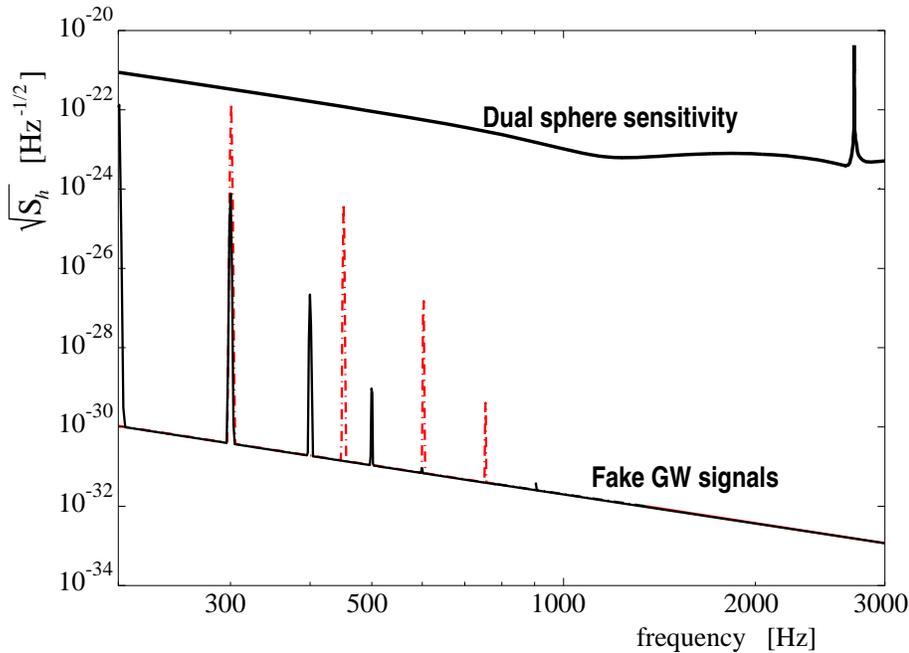}
\caption{GW equivalent signal generated by an oscillating passing by
mass, and sensitivity curve of a \emph{dual sphere} GW detector as
considered in reference~\protect\cite{dual}. The black solid line
corresponds to a mass oscillating at 100 Hz, while the the red
dash-dotted line corresponds to a faster oscillator of 150 Hz.
The plot very clearly shows that the signal generated by these
sources is orders of magnitude away from being detectable.
\label{fig.4}}
\end{figure}

We now use the expansion~(\ref{eq.4-7}) to determine the equivalent
signal~(\ref{eq.4-4}). The result is plotted in figure~\ref{fig.4},
where a dual sphere sensitivity curve has been added to assess the
real effect of two instances of fake signals.

The graphic very clearly shows how far the effect is from causing a
problem for gravitational wave physics, even though rather exaggerated
data have been assumed in both of the instances plotted. These are
the figures used in the plot:

\begin{center}
\begin{tabular}{lcl}
{\sf Impact parameter:} & \hspace*{1 em} & $d\/$\,=\,50 metres \\
{\sf Horizontal speed:} & & $v\/$\,=\,40 m/s\ \ \ ($\sim$\,140 km/h) \\
{\sf Oscillating mass:} & & $m'\/$\,=\,100 kg \\
{\sf Oscillation amplitude:} & & $A\/$\,=\,40 cm
\end{tabular}
\end{center}
and the frequencies of oscillation are 100 Hz (6000 rpm) and 150 Hz
(9000 rpm) in each of the examples, respectively. The results shown
in the graph scale with the tabulated values as

\begin{equation}
 \tilde h_{\rm equiv}\propto\frac{m'A}{vd^3}\;,
 \label{eq.4-9}
\end{equation}
while it is also seen that narrow peaks happen at the harmonic
frequencies of 100 Hz or 150 Hz ---actually of $\alpha/2\pi$, as
discussed above.

\section{Concluding remarks
\label{sec.5}}

The calculations in this paper show that there is no practical
possibility that noise generated by nearby traffic, which exactly
fakes quadrupole GW noise, disturbs GW astronomy in the sensitivity
band of the spherical detector, even if this is chosen as the best
possible \emph{dual sphere} system.

While this is particularly clear in the presented plots, we must
still stress that those results actually constitute \emph{upper bounds}
on the fake signals, due to inequalities~(\ref{eq.23}) and~(\ref{eq.4-4}).
In addition, emphasis should also be put on the fact that most of
the \emph{background} signal we see in figure~\ref{fig.4} is due
to \emph{aliasing} of the out-of-band frequency components.

Altogether then, it appears that installation of a very sensitive
spherical GW detector in a site only a few hundred metres from a
heavy traffic road is definitely not going to represent a problem
for the system performance.

\ack{
JAL acknowledges support received from the Spanish Ministry of
Science, contract number BFM2000-0604. AM thanks Generalitat de
Catalunya and Universitat de Barcelona for a contract.}




\section*{References}

\begin{thebibliography}{99}

\bibitem{lobo} J.A.\ Lobo, {Phys Rev D} {\bf 52}, 591 (1995), and
	J.A.\ Lobo, Mon Not Roy Astr Soc {\bf 316}, 173 (2000). See
	also W.W.\ Johnson and S.M.\ Merkowitz, Phys Rev Lett {\bf 70},
	2367 (1993)

\bibitem{clo} E.\ Coccia, J.A.\ Lobo and J.A.\ Ortega, {Phys Rev D}
	{\bf 52} 3735 (1995)

\bibitem{dual} M.\ Cerdonio, L.\ Conti, J.A.\ Lobo, A.\ Ortolan,
	L.\ Taffarello and J.P.\ Zendri, {Phys Rev Lett} {\bf 87}
	031101 (2001)

\bibitem{as93} P.\ Astone {\it et al.}, {Phys Rev D} {\bf 47} 362 (1993)

\bibitem{alle} W.O.\ Hamilton {\it et al.}, in \emph{First Edoardo Amaldi
	Conference on Gravitational Waves}, edited by E.\ Coccia, G.\ Pizzella
	F.\ and Ronga, World Scientific Publishing Co., Singapore (1995)

\bibitem{niobe} I.S.\ Heng, D.G.\ Blair, E.N.\ Ivanov and M.E.\ Tobar,
	{Phys Lett A} {\bf 218} 90 (1996)

\bibitem{naut} P.\ Astone {\it et al.}, {Astropart Phys} {\bf 7} 231 (1997)

\bibitem{aur} M.\ Cerdonio {\it et al.}, {Class Quant Grav} {\bf 14}
	1491 (1997)

\bibitem{chi2} L.\ Baggio {\it et al.}, {Phys Rev D} {\bf 61} 102001 (2000)

\bibitem{igec} Z.A.\ Allen {\it et al.}, {Phys Rev Lett} {\bf 85}
	5046 (2000)

\bibitem{mora} A.\ Morales, private communication (2002)

\bibitem{ll70} L.D.\ Landau, and E.M.\ Lifshitz, \emph{Theory of Elasticity},
        Pergamon (1970)

\bibitem{abra} M.\ Abramowitz, and I.A.\ Stegun, \emph{Handbook of
	Mathematical Functions}, Dover (1972)

\bibitem{fft} See e.g.\ R.\ Tolimieri, M.\ An, and C.\ Lu, \emph{Mathematics
	of Multidimensional Fourier Transform Algorithms}, Springer-Verlag
	(New York, 1993)
\end{thebibliography}
\end{document}